# Experimental verification of electro-refractive phase modulation in graphene

Muhammad Mohsin*,1, Daniel Neumaier[1], Daniel Schall[1], Martin Otto[1], Christopher Matheisen[1], Anna Lena Giesecke[1], Abhay A. Sagade[1] & Heinrich Kurz[1]

[1]Advanced Microelectronic Center Aachen (AMICA), Applied Micro and Optoelectronic (AMO) GmbH, Otto-Blumenthalstr. 25, 52074 Aachen, Germany
* Correspondence and requests for materials should be addressed to mohsin@amo.de

**Graphene has been considered as a promising material for opto-electronic devices, because of its tunable and wideband optical properties. In this work, we demonstrate electro-refractive phase modulation in graphene at wavelengths from 1530 to 1570 nm. By integrating a gated graphene layer in a silicon-waveguide based Mach-Zehnder interferometer, the key parameters of a phase modulator like change in effective refractive index, insertion loss and absorption change are extracted. These experimentally obtained values are well reproduced by simulations and design guidelines are provided to make graphene devices competitive to contemporary silicon based phase modulators for on-chip applications.**

In modern optical high-speed communication systems, phase shift keying is the standard method for data modulation[1,2]. While for fiber optical systems, phase modulators based e.g. on $LiNbO_3$ provide excellent performance, for integrated silicon (Si) photonic systems, there is not yet an ideal phase modulator available. The most widely used approach for realizing phase modulators in integrated Si photonic systems is based on p-n junctions (depletion or injection type) which provides high-speed performance enabling the generation of data rates up to 60 GBit/s [3]. However, the relatively weak electro-refractive effect in Si p-n junctions requires devices of mm-size to achieve a phase-shift of $\pi$ [4,5]. This is associated with a large footprint, high energy consumption and high insertion loss, and therefore alternatives are urgently needed.

Graphene, the two dimensional allotrope of carbon, is considered as a promising material for a wide range of photonic applications [6] because of its unique electro-optical properties[7]. Additionally, a wafer-scale CMOS compatible integration into a Si photonic platform is conceivable[8,9]. While calculations for graphene based electro-refractive modulators suggest significant advantages especially in terms of device footprint, operation speeds and energy consumption compared to Si based phase modulators[10-15], an experimental realization of such a device is still missing.

In this work, we report on the experimental demonstration of a broad-band electro-refractive phase modulator using graphene as active material. Key parameters of this device such as insertion loss, change in effective refractive index, and change in absorption are extracted from the experiments and simulations have been performed reproducing these values. The results are then compared to the state-of-the-art Si modulators using the typical figure of merits and an outline is given for realizing graphene modulators that can significantly outperform current Si based phase modulators.

We use a stack of graphene-oxide-graphene embedded into one arm of a Si waveguide based Mach-Zehnder interferometer (MZI), where graphene is located in the evanescent field of the Si waveguide. The chemical potential of the graphene is changed electro-statically by biasing the two graphene layers with respect to each other. Therefore, the effective refractive index of one MZI arm is changed which causes a shift in the transfer function of the MZI.

## Results

Fig. 1a illustrates schematic of our device. The MZI is realized on Si-on-insulator (SOI) platform with ridge waveguides (width = 375 nm, height = 220 nm) on top of 2 μm buried oxide (BOX). TE-polarized light was coupled in using grating couplers optimized for 1550 nm. The relative difference between the lengths of two MZI arms is 91 μm. To avoid cracking of monolayer graphene at the step edges of the waveguide, a layer of hydrogen silsesquioxane (HSQ) was first spin coated on the sample and thermally cured for 1h at 300°C[16-18]. The thickness of HSQ on top of waveguides is 85 nm. Subsequently, 10 nm of $Al_2O_3$ were deposited with atomic layer deposition (ALD) at 300 °C using $O_2$ plasma and trimethylaluminium (TMA) as precursors. Fig. 1b shows a cross-section SEM image of waveguide with combined 95nm of HSQ and $Al_2O_3$. A single layer of CVD grown graphene was transferred to the sample by the PMMA transfer method[17,19]. Afterwards, graphene was contacted with nickel and patterned to a length of 200 μm using optical lithography and oxygen plasma. After another atomic layer deposition of 90 nm $Al_2O_3$ at 150 °C using water vapors and TMA as precursors, a second CVD grown single layer graphene, which acts as counter electrode, was transferred, contacted, and patterned using the same methods described for the first layer. In order to passivate the second graphene layer, another 40 nm of $Al_2O_3$ were deposited. Finally, vias were etched through the $Al_2O_3$ layers wet chemically to access the two nickel electrodes. An optical image of the final device is shown in Fig. 1c.

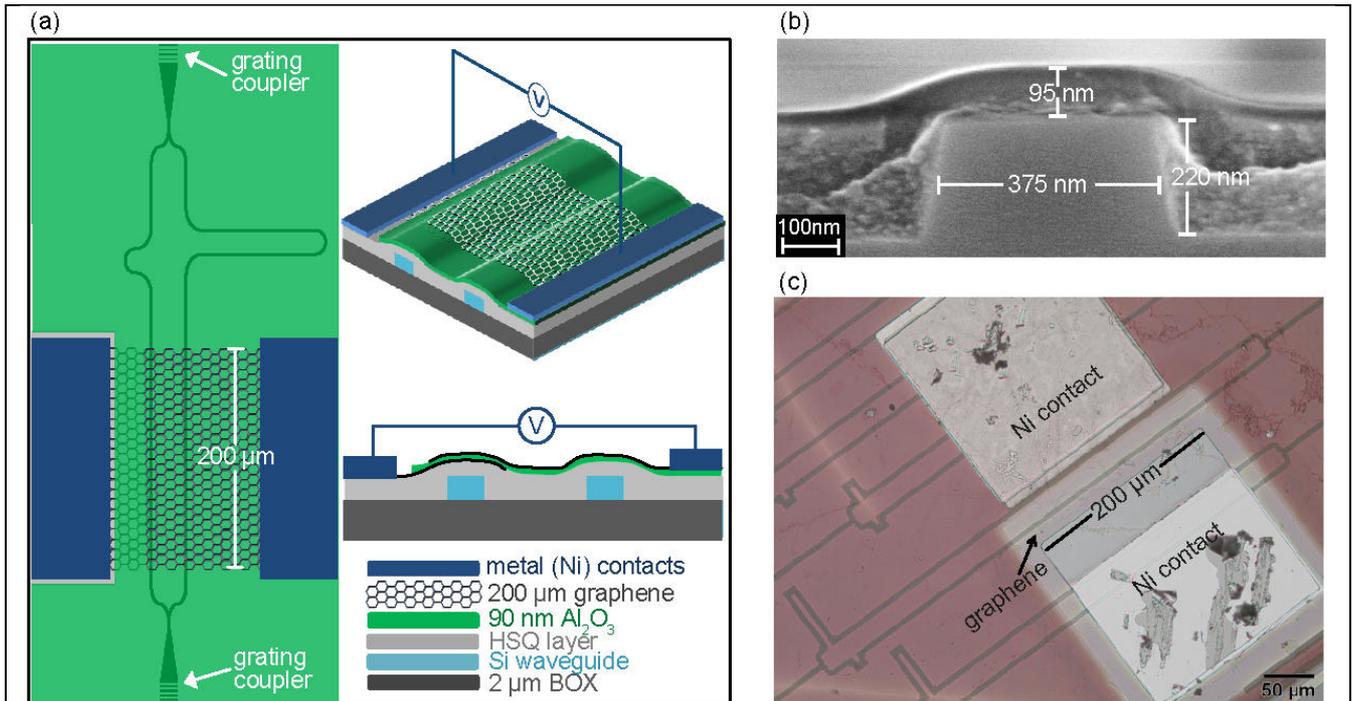

**Figure 1 Graphene based electro-refractive phase modulator.** (a) Schematic of the MZI used for determining Δn. Final layer of 40 nm $Al_2O_3$ is not shown for clarity. (b) Cross-section SEM image showing 95 nm (85 nm HSQ + 10 nm $Al_2O_3$) on top of MZI arm. (c) An optical image of final device.

All optical and electro-optical measurements were carried out in air at room temperature using a tunable continuous wave laser (1520-1620 nm) with 1 mW optical output power. To analyze the effect of each fabrication step on the transmission spectrum, the device was characterized at each stage of fabrication by measuring the transmitted optical power as a function of wavelength. Fig. 2a shows transmission spectra for three fabrication steps; i) with 85 nm HSQ and 10 nm $Al_2O_3$ on the sample (black spectrum), ii) after the first graphene layer was transferred, patterned, contacted, and covered by 90 nm $Al_2O_3$ (green spectrum) and iii) the final device (blue spectrum) with two graphene layers. These transmission spectra demonstrate clear interference pattern with a high extinction ratio of >15 dB for each mentioned step.

The grating couplers, y-splitters and Si waveguide account for an initial loss of ~15 dB as evident from the black spectrum in Fig. 2a. After contacting and patterning first graphene layer to 200 μm on one MZI arm and depositing 90 nm $Al_2O_3$ on top, the extinction ratio reduced from 22.5 dB to 17.5 dB (green spectrum in Fig. 2a). This reduction in extinction ratio is due to intrinsic absorption of graphene, which is only transferred to one arm of the MZI. From this reduction in extinction ratio, an intrinsic graphene absorption of ~2 dB (~0.01 dB/µm when normalized by graphene length) is extracted[20,21]. In the final device with two graphene layers (blue spectrum in Fig. 2a), the overall transmission is reduced. The extinction ratio, however, remained at the same level of >15 dB. The reduction of the overall transmission after each fabrication step is attributed to process induced contaminations. Apart from process induced contaminations, the dielectric layers, which get deposited on grating couplers, also reduce the coupling efficiency between optical fibers and grating couplers. The intrinsic graphene absorption, process induced contaminations and reduction in coupling efficiency between grating couplers and optic fiber are identified to be the main contributors to the overall device insertion loss.

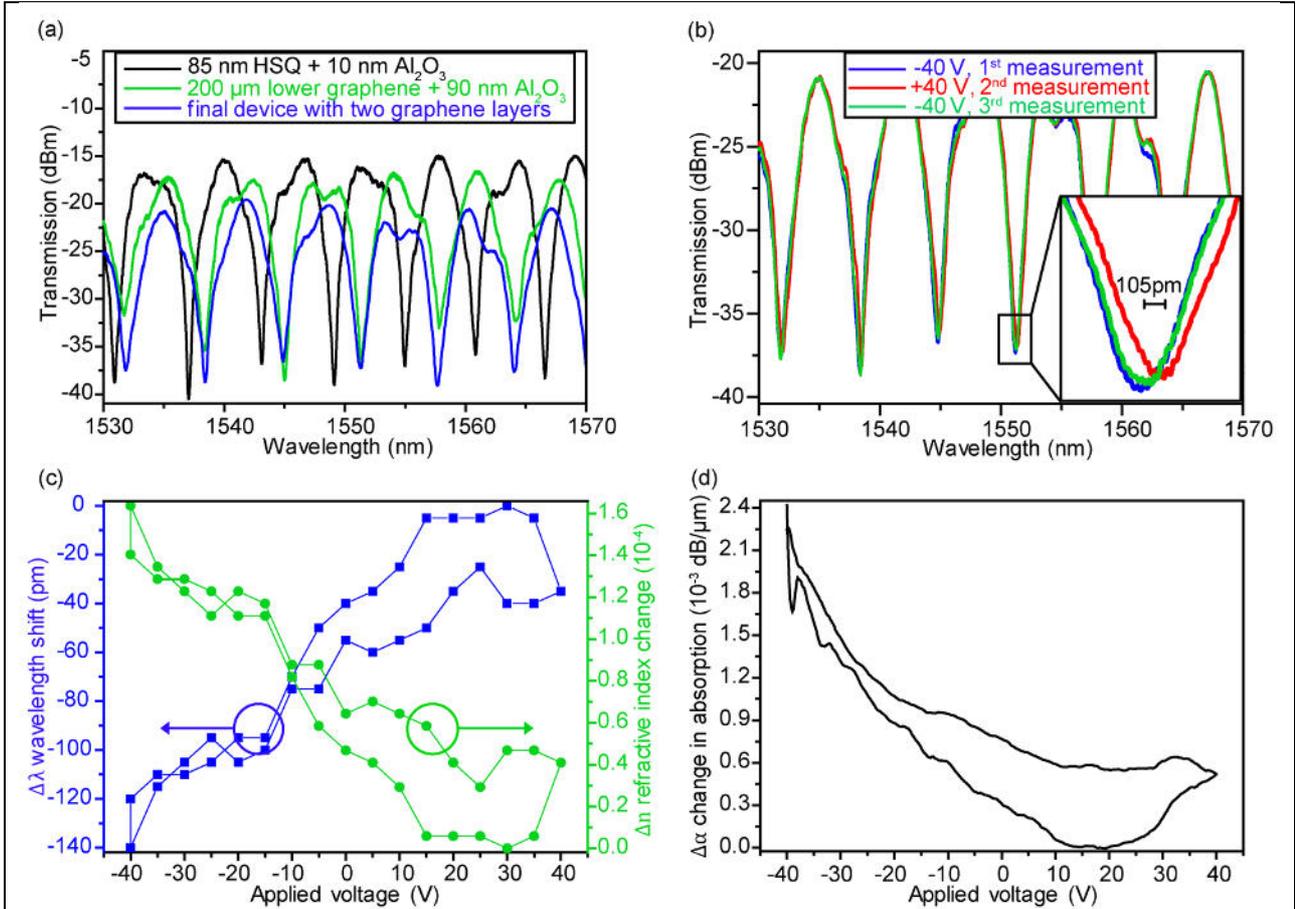

**Figure 2 Measurements performed under ambient conditions.** (a) Transmission spectra of device at different stages of fabrication. An intrinsic absorption of 0.01 dB/µm in lower graphene layer was estimated from the reduction in extinction ratio. (b) Applied voltages between two graphene layers cause a reproducible shift of transmission minimum, as is clear in inset. Only two distinct voltages have been plotted for clarity. (c) Wavelength shift Δλ as a function of applied voltage in steps of 5 V. The corresponding values of Δn, using eq. 1, are also plotted. (d) Change in absorption Δα as a function of applied voltage for the device. The hysteresis is mainly attributed to oxide grown with water process.

The optical transmitted power of the final device was measured as a function of the voltage applied between the two graphene layers from -40 V to +40 V and backwards, with the bottom graphene layer kept grounded. The applied voltage was relatively high because of the thick dielectric (90 nm) in between the two graphene layers. The thick dielectric was chosen in order to minimize light interaction of the second graphene layer, reducing the complexity of the system. Fig. 2b shows the transmission spectra for the two highest applied voltages (+40 V and -40 V). Inset depicts a clear and reproducible red shift of the minimum in transmission with increased voltages, demonstrating that the effective refractive index has been changed electro-statically.

By measuring the wavelength at the minimum of the transmission as a function of bias voltage between the two graphene layers, the change in refractive index (Δn) can be derived quantitatively using

$$\Delta n(V) = \frac{\lambda}{L}\left(\frac{\Delta\lambda(V)}{d}\right) \quad (1)$$

where L, d and Δλ are the graphene length (200 μm), the spacing between minimas (6.6 nm), and the wavelength shift with voltage V, respectively. Fig. 2c shows Δλ along with corresponding values of Δn. The maximum wavelength shift of 140 pm translates into a phase shift of π/20 induced by a change in effective refractive index of $1.5\times10^{-4}$.

In a MZI the change in absorbance Δα in one arm can be determined from the change in extinction ratio. As can be seen in Fig. 2b, an increase of the minimum transmission is observed at +40V, corresponding to an absorption change of 0.0028 dB/μm. However due to the relatively low Δα, the fitting of the transmission spectrum is associated with a high level of uncertainty. Therefore we converted our device to a pure electro-absorption modulator, by mechanically scratching one MZI arm (without the graphene modulator on top), which left an electro-absorption modulation as proposed in literature[22]. The light transmission of this electro-absorption modulator was measured for voltages from -40 V to +40 V. A maximum Δα = 0.0024 dB/μm was obtained as shown in Fig. 2d. The hysteretic behavior of the device characteristic is typical for graphene based field effect devices and has been related to $O_2/H_2O$ redox couples at the graphene/dielectric interface[23,24].

In addition to the experiments, simulations of the optical properties of the waveguide-graphene stack have been performed to get complementary information on the main optical parameters extracted in the experiments (absorption, Δn and Δα) and to explore the parameter space in terms of chemical potential and mobility. The simulations are based on the complex optical conductivity of graphene, which depends on the Fermi energy, the scattering rate and the temperature, and have been carried out using finite difference method in MATLAB[25]. Since the top graphene layer is more than 180 nm away from the waveguide and its effect on optical mode is found to be significantly smaller compared to the lower graphene layer, it is not considered in the simulations. The refractive indices of HSQ and $Al_2O_3$ are taken from literature[18,26]. As Fig. 3a illustrates, a stack of $SiO_2$-Si-HSQ-$Al_2O_3$-graphene-$Al_2O_3$ is considered with refractive indices of 1.44-3.48-1.38-1.64-$n_g$-1.64, respectively ($n_g$ being potential dependent refractive index of graphene) with TE mode propagating along the non-planar waveguide, which is an idealized situation of the stack used in the experiments. In the simulations, the complex optical conductivity of graphene (σ) is expressed as sum of intra-band and inter-band contributions which are determined using Kubo formalism given by[27,28],

$$\sigma_{intra} = \frac{ie^2 k_B T}{\pi\hbar^2(\omega + i2\Gamma)}\left(\frac{\mu_c}{k_B T} + 2\ln(e^{-\mu_c/k_B T} + 1)\right)$$

$$\sigma_{inter} = \frac{ie^2(\omega + i2\Gamma)}{\pi\hbar^2}\int_0^\infty \frac{f_d(-\xi) - f_d(\xi)}{(\omega + i2\Gamma)^2 - 4(\xi/\hbar)^2}d\xi$$

$$f_d(\xi) = \frac{1}{(e^{(\xi-u_c)/k_B T} + 1)}$$

where temperature (T), and Fermi velocity ($v_F$) are taken as 300 K, and $0.9\times10^6$ ms$^{-1}$, respectively[10,22]. Γ, $\mu_c$, ξ, e, ω, ℏ, $k_B$ and $f_d$ are the carrier scattering rate, chemical potential, energy, electron charge, radian frequency, reduced Planck's constant, Boltzmann constant and the Fermi-Dirac distribution, respectively. In the simulations, Γ is varied from 5e11 to 1e14 s$^{-1}$ in order to recognize its effect on the optical properties. These scattering rates correspond to charge carrier mobilities (μ) of 270 to 54000 cm$^2$/Vs at $\mu_c$= 0.3 eV (calculated using $\mu = (ev_F^2)/(\Gamma\mu_C)$), which are typically found in real devices. Since graphene was found to be p-doped for our device, the simulations have been discussed only for negative electro-chemical potentials here. However, the optical conductivity of graphene is symmetric for positive and negative

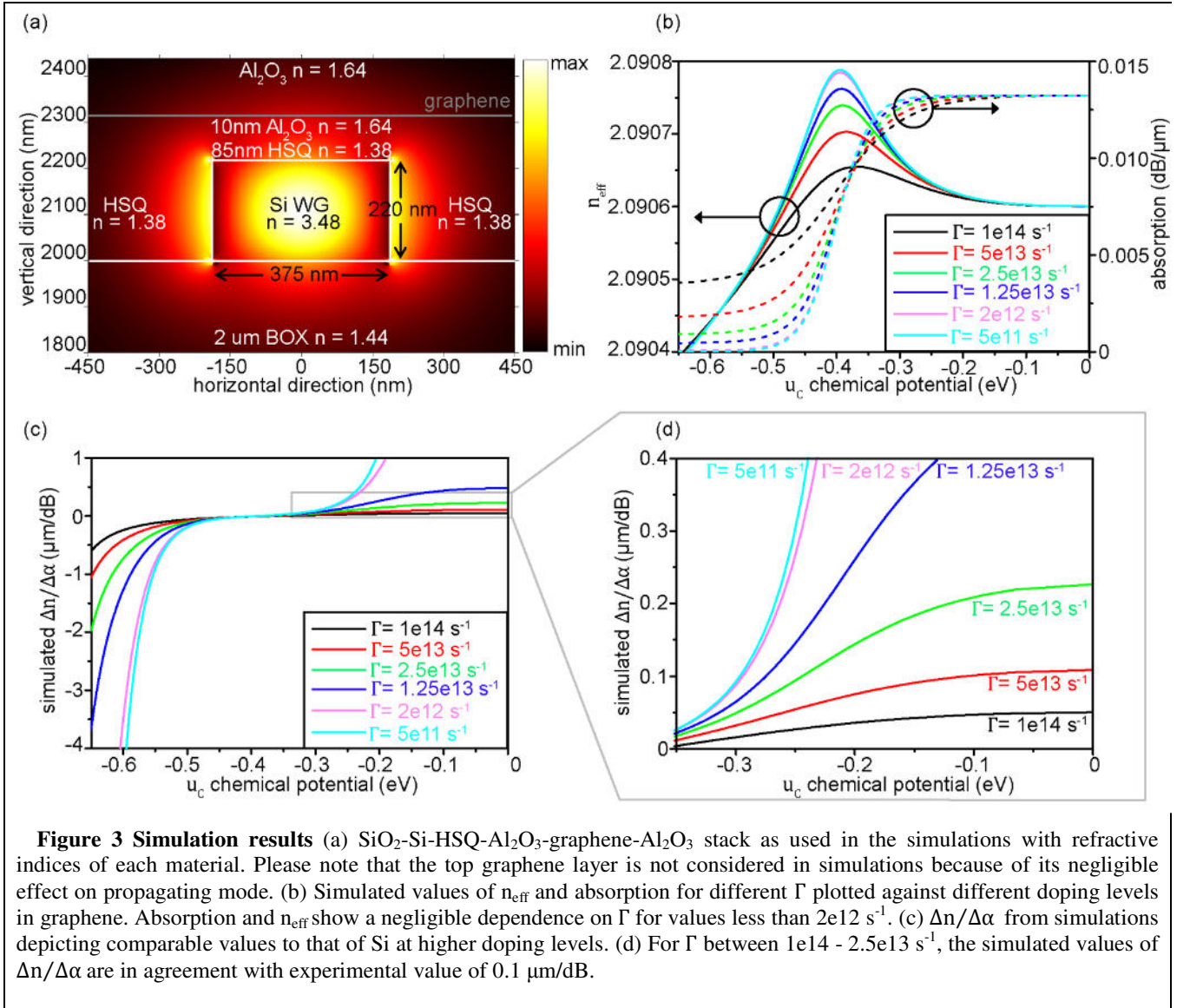

**Figure 3 Simulation results** (a) $SiO_2$-Si-HSQ-$Al_2O_3$-graphene-$Al_2O_3$ stack as used in the simulations with refractive indices of each material. Please note that the top graphene layer is not considered in simulations because of its negligible effect on propagating mode. (b) Simulated values of $n_{eff}$ and absorption for different $\Gamma$ plotted against different doping levels in graphene. Absorption and $n_{eff}$ show a negligible dependence on $\Gamma$ for values less than 2e12 s$^{-1}$. (c) $\Delta n/\Delta \alpha$ from simulations depicting comparable values to that of Si at higher doping levels. (d) For $\Gamma$ between 1e14 - 2.5e13 s$^{-1}$, the simulated values of $\Delta n/\Delta \alpha$ are in agreement with experimental value of 0.1 µm/dB.

electro-chemical potentials due to symmetric band structure in graphene[10,29].

The dielectric constant $\varepsilon_g$ (and hence refractive index $n_g$) of graphene is related to its optical conductivity by[28],

$$n_g = \sqrt{\varepsilon_g} = \sqrt{1 + \frac{i\sigma}{\omega t_g \varepsilon_0}}$$

where $t_g = 0.33$ nm is the thickness of graphene and $\varepsilon_0$ is the permittivity of free space. Using finite difference method, values of $n_{eff}$ and absorption have been calculated from eigen-solution of Maxwell equation[25],

$$\nabla \times (\epsilon^{-1} \times \nabla \times H) - \omega^2 \mu_0 H = 0$$
$$\nabla \times H = j\omega\epsilon E$$

where $\epsilon$ is dielectric permittivity tensor which takes into account refractive indices of $SiO_2$-Si-HSQ-$Al_2O_3$-graphene-$Al_2O_3$ stack. The eigen-solution of above Maxwell equation gives complex eigenvalues, with the real and imaginary parts representing $n_{eff}$ and absorption, respectively. The simulated values of $n_{eff}$ and absorption are plotted in Fig. 3b for different $\Gamma$. There graphene shows a simulated maximum intrinsic absorption of 0.013 dB/µm at $\mu_c$ = 0 eV, independent on $\Gamma$ and identical to experimentally obtained value of 0.01 dB/µm. At $\mu_c$

< -0.4 eV a strong dependency of absorption on $\Gamma$ is observed, as intra-band absorption becomes the dominating process there. In this regime, low $\Gamma$, corresponding to high carrier mobility, gives a lower absorption. In contrast to the absorption, $n_{eff}$ shows only a dependency on $\Gamma$ at the maximum value of $n_{eff}$ around $\mu_c \sim 0.4$ eV, and is effectively independent on $\Gamma$ for higher and lower $\mu_c$ as is clear from Fig. 3b.

**Discussions**

After the measured and simulated values of the graphene based phase modulator have been presented, a comparison with silicon based phase modulators can be given using different common figures of merit. One major figure of merit for a phase modulator is the product of modulator length L and drive voltage $V_\pi$ for a phase shift of $\pi$. In principle this product, termed as $V_\pi \cdot L$, should be as small as possible. For modulator realized here, a value of 30 V·cm is obtained, which is larger compared to Si based phase modulators, where typical values in the range of 0.5-15 V·cm are achieved[4,5]. However, such a large value is not unexpected in our case, as it is related to our device architecture, where a large distance between the lower graphene layer and the Si waveguide leads to relatively weak light interaction. In addition, the 90 nm thick dielectric between the two graphene layers causes weak electrostatic coupling. In our experiments, the main aim was to realize a proof-of-concept graphene based phase modulator using simplest fabrication steps. Optimizing the device architecture such as placing the lower graphene layer directly on top of the waveguide and reducing the dielectric thickness between the two graphene layers to 5 nm will result a $V_\pi \cdot L$ of ~0.2 V·cm, thus significantly outperforming state-of-the-art Si phase modulators. Using a technologically more challenging layout where the two graphene layers are placed in the middle of the waveguide[27,30], a $V_\pi \cdot L$ as low as 0.05 V·cm is realizable with 5 nm of $Al_2O_3$ as dielectric in between. These calculations are based on the scaling of the capacitance between the two graphene layers and by the evanescent fields at different locations, which were obtained from the simulations of the waveguide's mode profile and are consistent with previous simulations[11].

The insertion loss caused by intrinsic graphene absorption is 2 dB for the phase modulator realized in this work, while a phase shift of $\pi/20$ was achieved. This means that for a scaled device, which can perform a phase shift of $\pi$, the insertion loss would be ~40 dB, which is not acceptable for practical applications. These values are in good agreement to the simulations. In addition, the simulations suggest that at higher doping levels where $|\mu_c| > 0.5$ eV, the absorption is significantly reduced due to Pauli-blocking of the inter-band contribution. The insertion loss of a scaled phase modulator, which can perform a phase shift of $\pi$, would be only 2 dB at $\mu_c = -0.6$ eV and $\Gamma = 1.25e13$ s$^{-1}$ ($\mu$=1080 cm²/Vs at $\mu_c = -0.6$ eV). Here, lower $\Gamma$, i.e. higher carrier mobility, leads to an even lower insertion loss. This would be a significant improvement compared to Si MZI based phase modulators having an insertion loss of at least 4 dB[31].

Another figure of merit is $\Delta n/\Delta \alpha$, which defines the ratio of change in refractive index to the change in absorption. For our device an average value of 0.1 µm/dB is extracted from the experiments, which is a factor of 10 smaller compared to Si based modulators[32]. Again the experimental value is in agreement with simulations for a $\Gamma$ in the range of 2.5e13 to 5e13 s$^{-1}$ ($\mu_c$ is varied from 0 to approximately -0.35 eV in our experiments). These scattering rates correspond to a carrier mobility of 500-1000 cm²/Vs at $\mu_c = -0.3$eV, a mobility typically measured in reference devices using the same fabrication process. The low $\Delta n/\Delta \alpha$ of 0.1 µm/dB means that for obtaining a phase shift of $\pi$, the light intensity is changed by 10 dB, which is unacceptable for most applications demanding constant light intensity. Again our simulations suggests that a significant improvement can be expected either for lower $\Gamma$ or for higher doping levels where $|\mu_c| > 0.5$ eV. Under these conditions $\Delta n/\Delta \alpha$ can reach excellent values being larger than 1 µm/dB (see Fig. 3c,d).

In conclusion, an electro-refractive phase modulator, operating in wavelength range 1530-1570 nm, is realized experimentally using graphene as active material. Key parameters of the modulator such as absorption, $\Delta n$ and $\Delta \alpha$ have been extracted from the experiment and reproduced by simulations. While the parameters obtained from experiments are behind state-of-the-art Si based phase modulators, the simulations suggest that outstanding parameters for phase modulation can be achieved using graphene as active material. This requires first an enhanced interaction of the graphene with the waveguide mode and a stronger dielectric coupling between the two graphene layers in order to achieve competitive values of $V_\pi \cdot L$. Secondly, for achieving low insertion loss and high $\Delta n/\Delta \alpha$ values, $|\mu_c| > 0.5$ eV and a low scattering parameter (i.e. high carrier mobility) are required. Such high doping levels are realizable with molecular doping[33,34], while significantly higher mobility can be achieved using graphene encapsulated in hexagonal Boron Nitride[35]. As already shown in previous studies[14,15], high mobility also enable ultimate operation speeds. Therefore graphene offers an excellent basis for realizing ultra-fast phase modulators on a chip-integrated photonic platform.

**Acknowledgements**

This work was financially supported by European Commission under Contract No. 604391 (Project "GRAPHENE-FLAGSHIP"), Contract No. 285275 ("GRAFOL") and by the German Science Foundation DFG within the SPP 1459 Graphene (Project "GraTiS"). The Si waveguides were fabricated within the ePIXnet network of excellence.